# Proposal of Optimum Application Deployment Technology for Heterogeneous IaaS Cloud

Yoji Yamato[+]

Software Innovation Center, NTT Corporation, 3-9-11 Midori-cho, Musashino-shi, Tokyo, Japan

**Abstract.** Recently, cloud systems composed of heterogeneous hardware have been increased to utilize progressed hardware power. However, to program applications for heterogeneous hardware to achieve high performance needs much technical skill and is difficult for users. Therefore, to achieve high performance easily, this paper proposes a PaaS which analyzes application logics and offloads computations to GPU and FPGA automatically when users deploy applications to clouds.

## 1. Introduction

Recently, Infrastructure as a Service (IaaS) clouds have been progressed, and users can use computer resources or service components on demand (e.g., [1, 2]). Early cloud systems are composed of many PC-like servers. Hypervisors, such as Xen [3] or kernel-based virtual machines (KVMs) [4], virtualize these servers to achieve high computational performance using distributed processing technology, such as MapReduce [5].

However, recent cloud systems change to make the best use of recent advances in hardware power. For example, to use a large amount of core CPU power, some providers have started to provide baremetal servers which do not virtualize physical servers. Moreover, some cloud providers use special servers with strong graphic processing units (GPUs) to process graphic applications or special servers with field programmable gate arrays (FPGAs) to accelerate specific computation logics. For example, Microsoft's search engine Bing uses FPGAs to optimize search processing [6].

To use the recent advances in hardware power, users can benefit from high performance of their applications. However, to achieve this, users need to program appropriate applications for heterogeneous hardware and have much technical skill. Therefore, our objective is to enable users to achieve high performances easily. For this objective, cloud PaaS analyzes application logics and offloads computations to GPU and FPGA automatically when users deploy applications. The author previously proposed a Platform as a Service (PaaS) to select appropriate provisioning type of baremetal, container or virtual machine based on user requests [7]. In this paper, we investigate an element technology to offload part logics of applications to GPU and FPGA.

The rest of this paper is organized as follows. In Section 2, we review and clarify existing technologies problems. In Section 3, we propose a method of optimum application deployment for heterogeneous IaaS cloud. In Section 4, we conclude this paper.

## 2. Problems of Existing Technologies

Recently, GPU programming, such as the compute unified device architecture (CUDA) [8], that involves GPU computational power not only for graphics processing has become popular. Furthermore, to program

---

[+] Corresponding author. Tel.: +81 422 59 4395; fax: +81 422 59 2699.
*E-mail address*: yamato.yoji@lab.ntt.co.jp.

without walls between the CPU and GPU, the heterogeneous system architecture (HSA) [9], which allows shared memory access from the CPU and GPU and reduces communication latency between them, has been extensively discussed.

For heterogeneous programming, it is general to add and specify a code line to direct specified hardware processing. PGI Accelerator Compilers with OpenACC Directives [10] can compile C/C++/Fortran codes with OpenACC directives and deploy execution binary to run on GPU and CPU. OpenACC directives indicate parallel processing sections, then PGI compiler creates execution binary for GPU and CPU. Aparapi (A PARallel API) of Java [11] is API to call GPGPU (General Purpose GPU) from Java. Specifying this API, Java byte code is compiled to OpenCL and run when it is executed.

To control FPGA, development tools of OpenCL for FPGA are provided by Altera and Xilinx. For example, Altera SDK for OpenCL [12] is composed of OpenCL C Compiler and OpenCL Runtime Library. OpenCL C Compiler compiles OpenCL C codes to FPGA bit stream and configures FPGA logic, OpenCL Runtime Library controls FPGA from applications on CPU using libraries of OpenCL API. Programmers can describe FPGA logic and control by OpenCL, then configured logic can be offloaded to specified FPGA.

However, these technologies have two problems. A) General language codes of C, C++, Java need directives such as OpenACC or language extension such as Open CL C. If we would like to achieve high performance, timing to specify directives is very important and much technical knowledge is needed. B) There is no PaaS to utilize CPU/GPU/FPGA appropriately in clouds and users need to design how much GPU instances are needed.

The author previously proposed a PaaS to provide services based on user requests [7], [13], [14]. The work of [7] can provision baremetal, container or virtual machine appropriately, thus enhancing [7] idea, we can provide PaaS to select CPU/GPU/FPGA and can partly solve B). This paper targets to solve A) by an element technology to utilize GPU and FPGA from general language applications. Complex applications such as synchronous execution of FPGA and GPU are out of scope of this paper.

## 3. Proposal of Optimum Application Deployment Technology for Heterogeneous IAAS

In this section, we propose a cloud provider PaaS with optimum application deployment technology. Our proposed technology involves a PaaS, an IaaS controller, such as OpenStack [15], heterogeneous cloud hardware, and a code patterns database (DB). The figures describe OpenStack as an IaaS controller, but OpenStack is not a precondition of the proposed technology.

Fig. 1 shows system architecture and application deployment steps. There are 7 steps to deploy applications.

1. Users specify applications to deploy on clouds to PaaS system. Users need to upload source codes of applications to PaaS.

2. PaaS analyzes application source codes, compares codes to code patterns DB and detects similar code patterns. Here, code patterns DB retains codes which are offloadable to GPU and FPGA and corresponding OpenCL patterns. To detect similar codes, we use similar code detection tools such as CCFinderX [16]. Similar code detection tools can detect specified code patterns of FFT (Fast Fourier Transformation), encryption and decryption processing, graphic processing and so on from users' application codes. In these examples, FFT, encryption and decryption processing can be offloaded to FPGA with accelerated configurations of these processing, and graphic processing can be offloaded to GPU.

3. PaaS extracts OpenCL language codes for offloadable processing to GPU and FPGA detected in step 2. OpenCL language is major language for heterogeneous programming and describes processing which run on FPGA/GPU.

4. PaaS sends a provisioning request of creating run environment for applications to OpenStack. For example, when we need GPU, containers are provisioned on GPU servers because VMs cannot sufficiently control GPUs. And when we need FPGA, servers with FPGA board are provisioned by baremetal provisioning such as Ironic. Basically PaaS selects pre-configured FPGA server for specified logics such as FFT from multiple FPGA servers. However, there is no desired configuration of FPGA, PaaS may provision non-configured FPGA server and customized configuration may be done before actual applications run.

5. OpenStack creates computer resources for specified applications. Note that if applications need to create not only one compute server but several resources, such as virtual routers and storage, PaaS sends templates that describe the environment structures by JavaScript Object Notation (JSON) and provisions them at once by OpenStack Heat [17] or other orchestration/composition technology (e.g., [18]).

6. PaaS deploys application execution binary to provisioned servers. When PaaS deploys applications, existing tools of each vendor such as Altera SDK for OpenCL can be used.

7. PaaS returns application deployment results information of which servers are used for the processing and how much is the cost of server usage and so on. If users agree the deployment results, users start to use applications and usage fees are also started to charge. If users disagree the results, PaaS deletes resources by OpenStack Heat stack-delete API. After resources deletion, users may re-deploy. And if users would like to reconfigure FPGA, users specify FPGA configuration in this step 7 timing.

By these processing steps, users can deploy general language codes applications to heterogeneous cloud with extracting offloadable logics and OpenCL codes automatically. There is a merit for users to achieve high performance without program knowledge of GPU and FPGA.

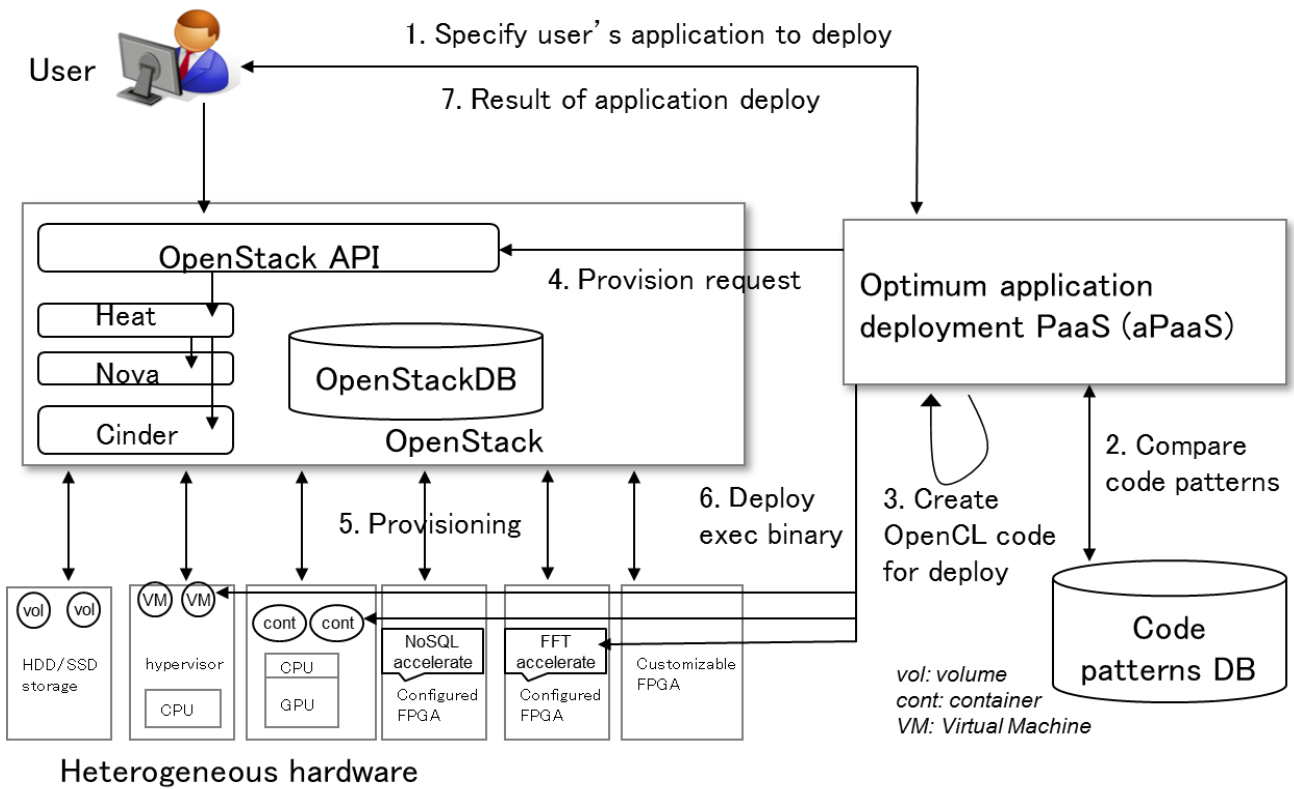

Fig. 1: Optimum application deployment steps for heterogeneous IaaS cloud.

## 4. Summary

This paper proposed a PaaS to offload application logics to GPU and FPGA automatically when users deploy applications to clouds. Proposed PaaS analyzed source codes, detected offloadable logics to GPU and FPGA using similar code detection technology and predefined code patterns, created OpenCL codes and deployed them. This can enable high performance applications easily for users. In the future, we will verify the proposed technology performance and validity for general application codes. We also study how to deploy complex applications in the future.